\documentclass{icrc}
\usepackage{times}
\usepackage{graphicx} 
\newcommand{\lesssim}{\lower.5ex\hbox{$\; \buildrel < \over\sim \;$}}
\newcommand{\gtrsim}{\lower.5ex\hbox{$\; \buildrel > \over\sim \;$}}

\newcommand{\gp}{\gamma^\prime}
\newcommand{\Dp}{\Delta^\prime}

\newcommand{\pp}{p^\prime}

\newcommand{\vp}{v^\prime}

\newcommand{\tp}{t^\prime}

\newcommand{\Ep}{E^\prime}
\newcommand{\qpa}{q-$\parallel$~}
\newcommand{\qpe}{q-$\perp$~}
\newcommand{\psim}{\lower.5ex\hbox{$\; \buildrel \propto \over\sim \;$}}

\begin{document}

\title{Maximum Particle Energies by Fermi Acceleration and the 
Origin of Cosmic Rays above the Knee}
\author[1]{C. D. Dermer}
\affil[1]{Code 7653, Naval Research Laboratory, Washington, DC 20375-5352 USA}

\correspondence{dermer@gamma.nrl.navy.mil}

\firstpage{1}
\pubyear{2001}


\maketitle

\begin{abstract}
We derive the maximum accelerated particle energy from first-order and
second-order Fermi acceleration at nonrelativistic and relativistic
shocks for explosions taking place in a uniform surrounding medium.
Second-order stochastic processes in relativistic flows are shown to
be capable of accelerating cosmic rays to ultra-high energies. Cosmic
rays above the knee of the cosmic ray spectrum can be accelerated by the
second-order Fermi mechanism in relativistic flows, such as those
occuring in gamma-ray bursts and unusual supernovae like SN 1998bw.
\end{abstract}

\section{Introduction}

Although there is a consensus that cosmic rays with energies less than
the knee energy at $\approx 3\times 10^{15}$ eV are accelerated by
supernova remnant shocks, the origin of higher energy cosmic rays is
unknown.  Proposed sources of cosmic rays above the knee include
galactic sources such as microquasars and pulsars, active galactic
nuclei \citep{ps92}, merger shocks in galaxy clusters \citep{krj96},
dormant AGNs \citep{lev01}, and gamma-ray bursts
(GRBs) \citep{mu96,dh01}. Proposed sources of ultra-high energy
($\gtrsim 10^{19}$ eV) cosmic rays (UHECRs) include the termination
shocks at the lobes of FRII jets \citep{rb93}, GRBs
\citep{wax95,vie95,der00}, and exotic particles, such as monopoles and
Z-bursts (for reviews, see \citet{nw00,wei01}).

Here we consider particle acceleration at nonrelativistic and
relativistic shocks. We show that the sources of cosmic rays near and
above the knee of the cosmic-ray spectrum could result from particle
acceleration at the relativistic shocks produced by the baryon-dilute
outflows of fireball transients and GRBs \citep{der00,dh01}.The
maximum energy that can be reached by a particle accelerated at a
shock due to constraints upon diffusive transport and available time
is derived, extending earlier treatments for nonrelativistic shocks
by, e.g., \citet{lc83}, \cite{dru83}, and
\citet{bar99}.

In \S 2, the maximum particle energy is derived for the extremes of
quasi-parallel (q-$\parallel$) and quasi-perpendicular (q-$\perp$)
shocks \citep{jok87,bar99}. We find that the maximum particle energy is not
greatly increased in the latter case due to a geometric constraint
that applies in cases where a large scale toroidal magnetic field is
absent. In \S 3, we consider the highest energy that particles can be
accelerated by shock Fermi processes at a relativistic external shock,
following \citet{ga99}. In \S 4 we show that the highest energies
that can be reached by Fermi processes in supernova or GRB-type
explosions result from second-order gyroresonant stochastic
acceleration in the shocked fluid of a relativistic blast wave
\citep{wax95,rm98,sd00,dh01}.  This maximum particle energy $E$ can exceed
$\approx 10^{20}$ eV in GRB blast waves for reasonable parameter
values. Cosmic rays with energies above the knee of the cosmic
ray spectrum is proposed to originate from second-order Fermi acceleration 
at relativistic shocks (\S 5).

\section{Nonrelativistic Shocks}

We treat first-order Fermi acceleration of relativistic nonthermal
particles with $v^\prime \approx c$. Our notation is that $\beta(x) c
= \beta c$ is the flow speed, and the blast-wave evolution is
described by momentum $P = \beta\Gamma$.  Primes refer to quantities
in the comoving shock frame. In the primed frame stationary with
respect to the shock, the upstream ($-$) flow approaches with speed
$u_- = \beta_-c = 4\beta c/3$ and the downstream (+) flow recedes with
speed $u_+ = \beta_+c = \beta c/3$. The quantity $u = u_--u_+ = \beta
c$ is the speed of the shocked fluid and $u_-$ is the speed of the
shock as measured in the stationary upstream frame.

Let $\pp = \sqrt{\gamma^{\prime 2}-1}$ represent the dimensionless
momentum of a particle with Lorentz factor $\gp$ in the comoving shock
frame. Then $\dot p^\prime_{\rm FI} \cong \Delta \pp/ t_{cyc}$, where
the cycle time $t_{cyc}$ is given by the diffusive properties of the
upstream and downstream regions. The average change in particle
momentum over a complete cycle for relativistic test particles with
$\pp \gg 1$ is $\Delta \pp = 4\beta\pp/3$, when $\beta\ll 1$
\citep{sch84,gai90}, provided that there are scatterers in the
upstream and downstream flow to isotropize the particles. In a
one-dimensional flow,
\begin{equation}
t_{cyc} = {4\over v^\prime}\;({\kappa_-\over u_-}+{\kappa_+\over
u_+})\; = {4\over v^\prime u_- } (\kappa_-+\chi \kappa_+),
\label{tcyc}
\end{equation}
where $\chi = u_-/u_+ $ is the compression ratio and the 
diffusion coefficient $\kappa_\pm = \lambda_\pm \vp/3 = \eta_\pm
r_{\rm L}\vp/3 = \eta_\pm r_{{\rm L}\pm}^o\pp\vp/3$, where $r_{{\rm
L}\pm} = r_{{\rm L}\pm}^o \pp = mc^2\pp /qB_\pm$ is the Larmor radius
for a particle of mass $m$ and charge $q$. Here we write diffusion
coefficients in terms of the parameters $\eta_\pm$ that give the
particle mean-free-paths scaled to the values implied by the Bohm
diffusion limit evaluated for the local magnetic field $B$.

In \qpa shocks, $B_-/B_+ \cong 1$. Hence
\begin{equation}
\dot p^\prime_{\parallel} \cong
 {u_-\beta\over r_{{\rm L}-}^o(\eta_-+\chi {B_-\over B_+}\eta_+)}\;
\lesssim  {u^2\chi\over c r_{{\rm L}-}^o (\chi^2 -1)}\,
 \equiv \dot p^\prime_{\parallel ,max}.\;
\label{qpa}
\end{equation}
The last relation defines the Bohm-diffusion-limit maximum
acceleration rate for \qpa shocks, assuming that the Bohm diffusion
limit sets a lower limit to the diffusivity of a medium.

In \qpe shocks, a scattering event shifts a particle in the $\hat x$
direction by a mean distance $r_{\rm L}$; therefore $\lambda
\rightarrow r_{\rm L}$. The effective particle drift speed is reduced
by a factor $\eta^{-1}$ because particles are confined within the
gyro-radius size scale $r_{\rm L}$. Hence $\kappa_\perp \cong r_{\rm
L}\vp/ 3\eta$ or $\kappa_\perp \cong \kappa_\parallel/(1+\eta^2)$
\citep{jok87}, and
\begin{equation}
\kappa_{\perp \pm}
= {\eta_\pm \over 1+\eta_\pm^2}\;{r_{{\rm L}\pm}\vp\over 3}\;;\;
r_{{\rm L}+}^o = {B_-\over B_+}r_{{\rm L}-}^o = {1\over\chi}r_{{\rm
L}-}^o\;,
\label{kperp} 
\end{equation}
implying
\begin{equation}
\dot \pp_\perp \lesssim  {u^2\over c r_{{\rm L}-}^o}\;{\chi\eta\over \chi^2 - 1}\; \equiv \dot p^\prime_{\perp ,max}\;,
\label{pperp}
\end{equation}
where $\eta = \min (\eta_-,\eta_+)$.

According to \citet{jok87}, to maintain an isotropic particle
distribution requires $\eta \lesssim
\vp/u \sim 1/\beta$ for relativistic particles. If shock
drift takes particles to regions of smaller obliquenesses, then the
acceleration rate is reduced \citep{jok87,bar99}. Particles will drift
along a \qpe shock to scatter in a region of smaller obliqueness when
$\eta r_{\rm L}-/ x \lesssim \theta_{1/2}\;,$ where $\theta_{1/2}$ is
a characteristic angle over which the shock obliquity changes. If the
surrounding magnetic field is homogeneous, $\eta_- \lesssim $
$\min[\beta^{-1},\sqrt{x/r_{{\rm L}-}}]$. The condition $\eta \lesssim
\sqrt{x/r_{{\rm L}-}}$ implies that $\eta \lesssim \beta^{-1/3}$.

The energy gain rates are more conveniently written in terms of the
differential distance $dx = \beta_{s}\Gamma_{s}c d\tp \rightarrow
\beta_{s}cd\tp = u_- d\tp \cong u_- dt$ that the shock advances during
the differential time element $dt$, as measured in the stationary
frame of the external medium. From the above results, we find that the
maximum energy gains per unit distance for quasi-parallel and
quasi-perpendicular nonrelativistic shocks are, respectively,
\begin{equation}
{d\Ep\over dx}|_{\parallel,max} \simeq \beta q B_-\;\;;\;\;
\label{dEpdxa}
{d\Ep\over dx}|_{\perp,max} \simeq \beta^{2/3} q B_-\;\;.\;\;
\label{dEpdxb}
\end{equation}

The maximum particle energy in nonrelativistic supernova shocks is
derived by considering a uniform medium with density $n_0$
that carries a homogeneous external magnetic field with intensity $B_-
= B_0 = 10^{-6} B_{\mu G}$ G. Shock-wave evolution is assumed to
follow the behavior of the coasting and adiabatic Sedov solutions. The
evolution of the blast wave momentum $P(x)=\beta(x)\Gamma(x)$ in the
adiabatic (Sedov) regime is described by the function
\begin{equation}
P(x) = {P_0\over \sqrt{1+(x/x_d)^3}}\cong 
 \cases{P_0 \; ,& $x\ll x_d$ \cr\cr
        P_0 ({x\over x_d})^{-3/2}\; , & $x_d \ll x $ \cr}\;
\label{P(x)}
\end{equation}
\citep{dh01}, where $P_0 = \sqrt{\Gamma_0^2-1} = \beta_0\Gamma_0$ is the initial blast-wave Lorentz factor that defines the baryon loading. The deceleration radius
\begin{equation}
x_d \equiv [ {3 (\partial E_0/\partial \Omega)\over \Gamma_0^2 m_p c^2
n_0}]^{1/3}\cong 2.1 ({m_\odot\over \Gamma_0^2 n_0})^{1/3}\;\rm{pc}\;
\label{x_d}
\end{equation}
\citep{mr93} is defined in terms of the directional energy release $m_\odot$ of
the explosion, expressed in units of Solar rest-mass energy per $4\pi$
sr. Thus $m_\odot = 1 \Leftrightarrow \partial E_0/\partial \Omega =
M_\odot c^2/4\pi = 1.4\times 10^{53}$ ergs sr$^{-1}$.  For a
nonrelativistic explosion, $x_d \cong \ell_{\rm S}$, the Sedov length
\citep{snp96}, where $ \ell_{\rm S} \equiv \Gamma_0^{2/3} x_d$.

The maximum particle energy for a particle accelerated from rest
starting near the location of the explosion is found by integrating
the energy gain rate per unit distance over the evolution of the blast
wave radius. For \qpa and \qpe nonrelativistic shocks, we obtain
\begin{equation}
E_{\parallel,max} \;\simeq \; 6\times 10^{15} Z \beta_0 ({m_\odot\over n_0})^{1/3} B_{\mu G}\;{\rm eV} \; 
\label{Epar}
\end{equation}
and
\begin{equation}
E_{\perp,max} \;\simeq\;   10^{16} Z \beta_0^{2/3} ({m_\odot\over n_0})^{1/3} B_{\mu G}\;{\rm eV} \; ,
\label{Eperp}
\end{equation}
respectively, where $Ze$ is the particle charge.

\section{Relativistic Shocks}

It is simplest to work in the frame of the upstream medium to
calculate the maximum energy of particles accelerated by a
relativistic external shock. In the first cycle, a particle increases
its energy by a factor $\Gamma^2$ but, due to kinematics of escape and
capture, the particle increases its energy by only a factor $\approx
2$ in successive cycles, as shown by \citet{ga99}. This is because a
particle is captured by the advancing relativistic shock when the
particle is deflected by an angle $\theta_d \sim 1/\Gamma_s$, so that
the upstream cycle time is $t_u \sim (\Gamma_s \omega_{\perp
-})^{-1}$. Here the particle gyration frequency in the upstream medium
is $\omega_{\perp -} = v/ r_{{\rm L}-}$, where $v$ is the particle
speed and the particle deflection is determined by $B_{\perp -}$, the
transverse component of the magnetic field in the upstream region. As
a result, the nonthermal particle distribution function at a
relativistic shock is highly anisotropic \citep{ks87}. As shown by
\citet{ga99}, the cycle time is dominated by the upstream transit time,

Because the energy increases by a factor $\cong 2$ during a single
cycle, the momentum gain rate following the first cycle in
relativistic shock acceleration is
\begin{equation}
\dot p_{rel} \simeq {2p\over t_u} \simeq {2cqB_{\perp -}\Gamma_s \over mc^2}\;,
\label{prel}
\end{equation}
provided $\Gamma_s \gg 1$.  If the particles are captured in the first
cycle with Lorentz factor $\bar \gamma$, then the energy from the
first cycle of Fermi acceleration in the stationary frame is $\approx
\Gamma^2\bar\gamma m c^2$ \citep{vie95}. After integrating the energy
gain rate, we find that the maximum energy that can be achieved by
relativistic first-order shock acceleration at an external shock is
therefore $$E_{rel,max} \simeq \bar\gamma\Gamma^2(x_0)mc^2 + 3\times
2^{3/2} q B_{\perp -} x_d \Gamma_0 $$
\begin{equation}
 \simeq [8\times 10^{13}\bar\gamma\Gamma_{300}^2 A + 10^{17} Z
 B_{\mu {\rm G}}({m_\odot\Gamma_{300}\over n_0})^{1/3} ]\;\;{\rm eV}\;,
\label{Erelmax}
\end{equation} 
where$Am_p$ is the particle mass, and $\Gamma= 300\Gamma_{300}$. From
equation (\ref{Erelmax}), we see that it is difficult to accelerate
particles to energies larger than a factor $\sim
2^{1/3}\Gamma_0^{1/3}$ over $E_{\perp, max}$, given by equation
(\ref{Eperp}), unless a pre-existing energetic particle distribution
is found in the vicinity of the explosion.

\section{Gyroresonant Stochastic Acceleration in Shocks}

We derive the particle acceleration rate through stochastic
gyroresonant processes from the expression $\dot p^\prime_{\rm FII}
\cong \Delta \pp/ t_{iso}$, where $t_{iso}$ is the pitch-angle
isotropization time scale in the comoving fluid frame. For
relativistic hard sphere scatterers, the fractional change in momentum
over time period $t_{iso}$ is $ \Delta p^\prime/\pp = 4 p_A^2/3$
\citep{gai90}. The quantity $p_A$c, which represents
 the dimensionless momentum of
the scattering centers, reduces to the Alfv\'en speed in the weakly
turbulent quasilinear regime. For relativistic shocks, the maximum
value that $p_A^2$ can take is given by $p_A^2 \cong 2e_B/3$, as can
be shown by considering the relativistic shock jump conditions
\citep{bm76}. The downstream
magnetic field $B_+ \cong \max (\chi\Gamma B_{\perp -},B_*
\sqrt{\Gamma^2 -\Gamma}~)$, where $ B_* \equiv (8\pi \chi n_0 m_p c^2
e_B)^{1/2} \cong 0.39 ({\chi\over 4} n_0 e_B)^{1/2}\;{\rm G}$. The
first term represents the compression of the upstream transverse
magnetic field, and the second term defines the downstream field in
terms of the equipartition field $B_{eq}({\rm G}) = 0.39 (e_B n_0)^{1/2}$
$\sqrt{\Gamma^2 -\Gamma}$ through the parameter $e_B$. For this
convention, $\chi = 4$ for strong shocks whether or not the shocks are
relativistic.

The particle pitch angle changes by $\delta B/B$ during one gyroperiod
$t_{gyr}= r_{{\rm L}+}/c$, where $B$ now refers to the mean downstream
magnetic field in the shocked fluid. Thus $t_{iso} \cong
t_{gyr}/(\delta B/B)^2$, noting that particles diffuse in pitch angle.
Pitch angle changes due to gyroresonant interactions of particles with
resonant plasma waves are described by the relation $(\delta B/B)^2
\approx \bar k W(\bar k)/U_B$, where $U_B = B^2/8\pi$, and the
turbulence spectrum $W(k) = W_0 (k/k_{min})^{-v}$ for $k_{min} \leq k
< k_{max}$. The index $v = 5/3$ for a Kolmogorov turbulence spectrum,
and $v = 3/2$ if the Kraichnan phenomenology is used. The resonant
wavenumber is normally assigned through the resonance condition
$\omega - k_\parallel v_\parallel = \ell \Omega/\gamma$, but here we
proceed by employing the simple resonance assumption $k\rightarrow
1/{r_{\rm L}+}$ \citep{bs87}. Assuming isotropy of forward and
backward-moving waves gives the normalization $W_0 = \xi U_B
(v-1)/2k_{min}$, where $\xi$ is the ratio of plasma turbulence to
magnetic field energy density. Hence $t_{iso}^{-1} \cong c\xi (v-1)
(r_{{\rm L}+} k_{min} )^{v-1}/(2 r_{{\rm L}+})$, and
\begin{equation}
\dot \pp_{\rm FII} \cong {2\over 3} p_A^2\; \xi (v-1)\; 
({c\over r_{{\rm L}+}^o})\;(r_{{\rm L}+}^o k_{min} \pp)^{v-1} \;
\label{dotpFII}
\end{equation}
(\citet{dml96}).

The term $r_{{\rm L}+}^o k_{min} \pp$ in the parentheses of equation
(\ref{dotpFII}) gives the comoving gyroradius in units of the inverse
of the smallest turbulence wavenumber $k_{min}$ found in the shocked
fluid. If $k_{min}\sim 1/\Dp$, where $\Dp$ is the blast wave width,
then the relation $r_{{\rm L}+}^o k_{min} \pp \lesssim 1$ is an
expression of the condition that the particle gyroradius is less than
the size scale of the system \citep{hil84}. When this condition is
satisfied, $ \dot \pp_{\rm FII,max}$ $\approx p_A^2 c/r_{{\rm L}+}^o$
$ \cong 2e_Bc/(3r_{{\rm L}+}^o)$.

By integrating the energy gain-rate for an adiabatic blast wave that
evolves according to equation (\ref{P(x)}), we obtain the maximum
energy-gain rates due to stochastic Fermi acceleration for
nonrelativistic ($\Gamma - 1 \ll 1$) and relativistic ($\Gamma \gg 1$)
shocks, given by
\begin{equation}
{d\Ep\over dx}|_{FII,NR} =
 {e_B \xi (v-1)\over 2^{3/2}} \;qB_* \beta^2
 \;({2^{1/2}\Ep\over q B_* f_\Delta x \beta})^{v-1}\;
\label{dEpdxFIINR}
\end{equation}
and 
\begin{equation}
{d\Ep\over dx}|_{FII,ER} = {2^{3/2}\over 9} e_B 
\xi (v-1) \;qB_* \;({\Ep\over q B_* f_\Delta x})^{v-1}\;,
\label{dEpdxFIIER}
\end{equation}
respectively. Here $f_\Delta$ sets the scale for the smallest
turbulence wave numbers in comparison with the inverse of the comoving
size scale $x/\Gamma$ of the blast wave width. From hydrodynamic
considerations, $f_\Delta = 1/12$.

For an adiabatic blast wave that explodes in a uniform surrounding medium, the
nonrelativistic expression (\ref{dEpdxFIINR}) can be integrated to
give the maximum energy
\begin{equation} 
E_{max,FII,NR} \cong  K_N\;qB_* f_\Delta x_d \beta_0 \;.
\label{Ex1NR}
\end{equation}
where 
$$K_N \equiv [{2^{v/2}\over 4} {e_B\xi\over f_\Delta}\;
 (v-1)(2-v) \beta_0 I_{2N}]^{1/(2-v)}\rightarrow $$
\begin{equation}
 \cases{2\times 10^{-7} \; [{e_B\xi \beta_{-2}(I_{2N}/6)
\over f_\Delta}]^3 \; ,& for $v = 5/3$  \cr\cr
        4\times 10^{-5} \; [{e_B\xi \beta_{-2}(I_{2N}/6)
\over f_\Delta}]^2 , & for $v=3/2 $  \cr}\;.
\label{K}
\end{equation}
Here $\beta_0 = 10^{-2}\beta_{-2}$, and $I_{2N}$ is an integral that
takes a maximum value $\simeq 6$.

The maximum energy of particles accelerated by second-order Fermi
acceleration in a nonrelativistic blast wave is defined by the
quantity
\begin{equation}
qB_* f_\Delta x_d \beta_0 \cong 8\times 10^{18} Z e_B^{1/2} 
(n_0)^{1/6}f_\Delta \beta_{-2} m_\odot^{1/3}\;{\rm eV}\;,\;
\label{qBfxGNR}
\end{equation} 
though the actual maximum energy is much less than this 
value when $\beta_0 \ll 1$, 
as can be seen from equation (\ref{K}).

The relativistic expression (\ref{dEpdxFIIER}) can be integrated to give
\begin{equation} 
E_{max,FII,ER} \cong  [{2^{3/2}\over 9} {e_B\xi (v-1)
\over f_\Delta}]^{\frac{1}{2-v}}\; 
qB_* f_\Delta x_d \Gamma_0 \;.
\label{Ex1}
\end{equation}
The maximum energy of particles
accelerated by second-order Fermi acceleration in a relativistic blast
wave occurs near the deceleration radius $x \approx x_d$. The basic
scaling for this maximum energy is given by the term
\begin{equation}
qB_* f_\Delta x_d \Gamma_0 \cong 7.7\times 10^{20}
 Ze_B^{1/2} n_0^{1/6}f_\Delta  (m_\odot \Gamma_0)^{1/3}\;{\rm eV}\;.\;
\label{qBfxG}
\end{equation}
Note the sensitive dependence of the maximum energy given
by equation (\ref{Ex1}) on the factor $ e_B \xi /f_\Delta$. If
$e_B\xi/f_\Delta$ $ > 1$ when the blast wave is near the deceleration
radius, then stochastic acceleration processes in relativistic blast
waves can in principle accelerate particles to ultra-high energies.


\section{Discussion}

The maximum particle energy that can be achieved through first-order
Fermi acceleration of a particle initially at rest, for an adiabatic
blast wave in a uniform surrounding medium, is given by
\begin{equation}
E_{max,1} \simeq 10^{16} Z B_{\mu {\rm G}}
 \beta_0^{2/3} ({m_\odot \Gamma_0\over n_0})^{1/3}\;\;{\rm eV}\;.
\label{Emax1}
\end{equation}
Here we use the \qpe result for nonrelativistic blast waves, which
gives the largest energies. The maximum particle energy that can be
achieved through second-order Fermi acceleration in a relativistic
blast wave is given by
\begin{equation}
E_{max,2} \simeq 8\times 10^{20} K_v Z e_B^{1/2}
 n_0^{1/6} f_\Delta  (m_\odot\Gamma_0)^{1/3}\;\;{\rm eV}\;,
\label{Emax2a}
\end{equation}
where $K_v = [2^{3/2} e_B \xi /9 f_\Delta]^{1/(2-v)}$. The strong
dependence of the maximum energy on $\beta_0$ makes second-order
processes unimportant in nonrelativistic flows when $\beta_0 \ll 0.1$.
As the flow speeds become marginally relativistic, first-order
processes may accelerate particles to sufficiently high energies that
second-order acceleration then starts to become important
\citep{sch84}. At relativistic shocks, the second-order process
can easily dominate.

Spectral models of GRBs within the external shock scenario
\citep{mr93,cd99} imply $\Gamma_{300} \sim 1$ and predict the
existence of clean and dirty fireballs which have not yet been
detected \citep{dcb99}.  Mildly relativistic outflows are deduced from
radio observations of the Type Ic SN 1998bw associated with GRB 980425
\citep{wei00,kul98}, so that hadronic cosmic ray acceleration to energies
above the knee of the cosmic ray spectrum is possible from Type Ic SNe.

First-order Fermi processes in relativistic flows can accelerate
particles to energies above the knee of the cosmic-ray spectrum. The
smooth cosmic ray spectrum between the knee and the ankle suggests
that the cosmic rays with energies above the knee originate from
sources that accelerate particles to energies $\gtrsim 10^{19}$ eV.
The second-order process in relativistic shocked fluids can
accelerate particles to ultra-high energies to form both the galactic
halo cosmic rays with energies between the knee and the ankle, as well
as the metagalactic UHECRs. The GeV cosmic rays detected near Earth
are probably a mixture of accelerated particles produced by many
sources, but the cosmic rays formed near and above the knee may be
produced by only one or a few energetic explosions \citep{ew00}. GRBs
are a prime candidate for the origin of these cosmic rays
\citep{mu96,dh01}.

\begin{acknowledgements}
This work is supported by the Office of Naval Research. Thanks are due to the referee of a related paper for correcting errors in an early version of this work.
\end{acknowledgements}

\clearpage

\end{document}